\documentclass{INTERSPEECH2023}

\interspeechcameraready 

\renewcommand{\footnotesize}{\fontsize{7pt}{2pt}\selectfont}
\usepackage[table,xcdraw]{xcolor}

\title{ASR and Emotional Speech: A Word-Level Investigation of the Mutual Impact of Speech and Emotion Recognition}
\name{Yuanchao Li$^{\dag}$\thanks{$^{\dag}$Equal contribution.}, Zeyu Zhao$^{\dag}$, Ond\v{r}ej Klejch, Peter Bell, Catherine Lai}
\address{Centre for Speech Technology Research, University of Edinburgh, UK}
\email{yuanchao.li@ed.ac.uk}

\begin{document}

\maketitle
 
\begin{abstract}
In Speech Emotion Recognition (SER), textual data is often used alongside audio signals to address their inherent variability. However, the reliance on human annotated text in most research hinders the development of practical SER systems. To overcome this challenge, we investigate how Automatic Speech Recognition (ASR) performs on emotional speech by analyzing the ASR performance on emotion corpora and examining the distribution of word errors and confidence scores in ASR transcripts to gain insight into how emotion affects ASR. We utilize four ASR systems, namely Kaldi ASR, wav2vec2, Conformer, and Whisper, and three corpora: IEMOCAP, MOSI, and MELD to ensure generalizability. Additionally, we conduct text-based SER on ASR transcripts with increasing word error rates to investigate how ASR affects SER. The objective of this study is to uncover the relationship and mutual impact of ASR and SER, in order to facilitate ASR adaptation to emotional speech and the use of SER in real world.
\end{abstract}


\section{Introduction}
Speech Emotion Recognition (SER) has rapidly developed over the past two decades, and the research on the corpora, features, algorithms, and training models has boomed \cite{schuller2018speech}. Emotion in speech is conveyed both by what is said and how it is said. The former is generally captured as text (i.e., speech transcripts). Recent work on SER consistently shows the benefits of incorporating both textual and acoustic features on benchmark corpora \cite{sebastian2019fusion, yoon2018multimodal}. Although some researchers have also attempted to translate this approach into the wild, e.g., in-car voice systems \cite{li2021feeling}, it is still rare to see SER applications in our daily lives. One of the major reasons is that the majority of SER research uses human annotation, i.e., gold-standard manual transcripts. In contrast, even for the `lab' emotion corpora (e.g., IEMOCAP), transcripts from a state-of-the-art Automatic Speech Recognition (ASR) system can result in Word Error Rates (WERs) higher than 35\%, depending on the emotion label \cite{li2022fusing}. This means that very few of the findings obtained in the lab can be replicated in the wild.

To bridge this gap, it is necessary to exploit the imperfect textual data generated by ASR for emotion recognition. While previous studies proposed using ASR transcripts for SER \cite{li2022fusing,sahu19_interspeech}, the effect of ASR on emotion and vice-versa is not always clear. For example, a high WER on emotional speech is generally assumed to be a result of the distortion relative to neutral speech \cite{fernandez2004computational}. However, \cite{li2022fusing} find neutral speech can in fact have a higher WER than angry speech, depending on utterance length. Thus, even though ASR is a relatively well studied area, key questions about how it can be used in SER remain unresolved, such as 1) \textit{how confident can we be of ASR on different emotions?} and 2) \textit{how is the SER performance affected by ASR errors?}  Thus, to make true progress in SER, we need to understand the interrelationship between ASR and SER.  


In this work, we look at these fundamental yet long-standing issues by exploring the mutual impact of ASR and emotional speech, with the objective of pushing SER research closer to realistic use scenarios. Specifically, we first take a deep look into the word distribution of emotion corpora and how they are misrecognized in speech recognition. We analyze WER variation to see how ASR performance is affected by different emotions and how this relates to word-level confidence scores. Finally, we produce several ASR transcripts with progressively increased WER to investigate how SER is affected by ASR performance. Our results demonstrate that: \textbf{1)} the ratio of words in terms of the Part of Speech (PoS) tag, affective scores, and utterance length can greatly affect ASR performance. \textbf{2)} Differences with respect to these characteristics in emotional speech lead to differences in ASR performance bias transcripts, and hence SER. \textbf{3)} The relationship between SER and ASR confidence scores requires careful analysis as their interpretation varies between ASR confidence with and without calibration. \textbf{4)} WER can greatly affect SER performance in a way that has implications for the use of real-life SER.

\section{Related Work}
Although the relationship between ASR and SER is an important topic, there is very limited work using ASR transcripts for SER.  \cite{schuller2009emotion} evaluated the emotion recognition performances with an ASR engine in the loop. Their experiments showed surprisingly low performance degradation ($<$3\%) with real ASR transcripts over ground-truth transcripts based emotion recognition. \cite{sahu19_interspeech} used two commercial ASR systems to generate transcripts for bimodal SER (audio + text), resulting in a relative loss of 4\% and 5.3\% in unweighted accuracy compared to ground-truth transcripts, respectively. \cite{li2022fusing} found that emotion classes with low WER typically have a low ratio of short utterances to long utterances, where the shorter the utterance, the higher the WER. They also showed that hierarchically fusing ASR hidden states and transcripts can achieve similar performance as ground-truth text. \cite{santoso2021speech} proposed the use of a confidence measure to adjust the importance weights in ASR transcripts according to the likelihood of a speech recognition error in each word. This mitigated the effects of ASR error on SER performance, achieving about 4\% accuracy increase compared to a baseline work without confidence measure.

These findings suggest it is possible to utilize ASR for SER, but they still have limitations. \cite{li2022fusing} only investigated utterance length but did not take other factors into consideration. \cite{sahu19_interspeech,schuller2009emotion,santoso2021speech} did not investigate individual emotions, though ASR may behave differently on them. Furthermore, none of these studies have been conducted thoroughly at the word level, and their findings cannot be generalized to other ASR models and corpora. Thus, the interrelationship between ASR and SER is still unclear, and gaps remain on foundational issues such as how ASR confidence varies with different emotions and how word distribution affects ASR performance. Therefore, we perform a word-level investigation on emotional corpora using multiple ASR models, conducting new analyses on ASR performance, confidence, and SER performance, towards understanding the mutual impact of speech and emotion recognition to best bring ASR into SER research.

\section{Emotion Corpora and ASR Models}
We use three corpora: IEMOCAP \cite{busso2008iemocap}, MOSI \cite{Zadeh2016} and MELD \cite{poria2019meld} to cover as many speech conditions as possible for generalizability. IEMOCAP (IEM) consists of five sessions of scripted and improvised dialogues in the research lab. We use four emotion classes: angry, happy (+excited), neutral, and sad, following previous research \cite{li2019improved}. MOSI has 2,199 monologue video clips from YouTube, annotated with sentiment scores in the range of [-3, 3]. MELD contains acted dialogue from the TV series \textit{Friends}. Each utterance has been labeled with one of seven emotion classes: anger, disgust, fear, joy, neutral, sadness, and surprise. For IEMOCAP and MELD, we removed utterances whose transcript is blank or whose audio file is too long due to mis-processing in corpora construction, bringing the total utterance numbers to 5,500 and 13,689, respectively.

For the ASR models, we use the Kaldi Librispeech ASR model\footnote{\url{https://kaldi-asr.org/models/m13}} (KLA), a self-supervised model: wav2vec2-base-960h\footnote{\url{https://huggingface.co/facebook/wav2vec2-base-960h}} (W2V2), a Conformer model\footnote{\url{https://github.com/espnet/espnet/blob/master/egs/librispeech/asr1/RESULTS.md}} (CONF) from ESPnet \cite{watanabe2018espnet}, and the medium model of Whisper\footnote{\url{https://openai.com/research/whisper}} (WHIS). KLA, W2V2, and CONF are pre-trained on Librispeech960, and WHIS is on 680,000 hours of multilingual and multitask data collected from the web. We use these four models for generalizability.

\section{Experiments and Analyses}
In this section, we analyze the relationship between ASR and emotion by conducting a series of experiments to identify sources of errors across corpora and word classes and quantifying the effect of WER on SER performance.

\subsection{WER Across Corpora} 

\begin{table}[ht!]
\centering
\caption{WER (\%) of the ASR models on the emotion corpora.}
\vspace{-5pt}
\label{wer}
\scalebox{0.95}{
\begin{tabular}{lccc}
\hline
 & \textbf{IEM} & \textbf{MOSI} & \textbf{MELD} \\ \hline
KLA & 36.8 & 40.9 & 58.5 \\ 
W2V2 & 32.7 & 35.4 & 57.8 \\ 
CONF & 27.1 & 30.1 & 52.1 \\ 
WHIS & 12.3 & 17.3 & 34.8 \\ \hline
\end{tabular}}
\end{table}

We compute the WERs of the four ASR models on the three emotion corpora as the basis for analysis, as shown in Table~\ref{wer}. Among the four models, WHIS shows the best performance on all the corpora, significantly outperforming the other three.
We see large performance differences among corpora, even though they are all US or UK English emotional speech. This could be due to the different recording settings, the noise that occurred, or the vocabulary that was said.



\subsection{PoS and Affective Score Effects on ASR}
\label{sec:word-class}
In addition to the well-known factors that affect ASR (e.g., recording quality, noise), we would like to know if word distribution also affects WER. To investigate this, we analyze the words in each corpus based on two aspects: Part-of-Speech (PoS) tag and affective score. In particular, we explore whether words of a particular class or with strong affective coloring have a strong impact on ASR performance.

\textbf{PoS.} We select seven classes of PoS tags: \textit{Noun, Verb, Adjective, Adverb, Wh-word, Function-word} from the Penn Treebank \cite{marcus1993building} and \textit{Stop-word} from the stopwords corpus of NLTK \cite{bird2006nltk}. The tagging was also conducted using NLTK.

\textbf{Affective score.} We use an affective words database that has nearly 14 thousand English words \cite{warriner2013norms} to refer the affective scores. Each word is rated in three affective dimensions: Valence (V), Arousal (A), and Dominance (D), ranging from 1 (weak) to 9 (strong). We divide the scores into three classes: low (1-3), medium (4-6), and high (7-9) according to their overall mean value, respectively.

We calculate three metrics to see how different classes of words affect the ASR performance. The Word Ratio (WR) is used to see the distribution of different classes of words in a corpus. The Error Ratio (ER) indicates if misrecognized words largely come from a specific class. The Class Error Rate (CER) measures how difficult words in a class are to recognize. We define them as follows.





\begin{equation}
\text{word ratio} = \dfrac{\text{word count per class}}{\text{total word count}}
\end{equation}

\begin{equation}
\text{error ratio} = \dfrac{\text{word error count per class}}{\text{total word error count}}
\end{equation}

\begin{equation}
\text{class error rate} = \dfrac{\text{word error count per class}}{\text{total word count per class}}
\end{equation}


The WR\footnote{Note that \textit{Stop} words can have multiple PoS tags as \textit{Stop} is not in the Penn Treebank, making the summation of the word ratios greater than 100\%.}, ER, and CER based on PoS tag and affective score are shown in Table~\ref{distribution-postag} and \ref{distribution-affect}, respectively. For brevity, we averaged the values over the four models. As the ER values from the four ASR models are close and the CER values are proportionally distributed, taking the average does not affect the general findings.

\begin{table}[ht!]
\caption{WR, ER, and CER (all in \%) based on PoS tag.}
\vspace{-5pt}
\centering
\label{distribution-postag}
\scalebox{0.775}{
\begin{tabular}{lrrrrrrrrr}
\hline
 & \multicolumn{3}{c}{\textbf{IEM}} & \multicolumn{3}{c}{\textbf{MOSI}} & \multicolumn{3}{c}{\textbf{MELD}} \\
 & WR & ER & CER & WR & ER & CER & WR & ER & CER \\ \hline
\textit{Noun} & 19.1 & 34.8 & \multicolumn{1}{r|}{33.3} & 21.7 & 36.8 & \multicolumn{1}{r|}{45.3} & 25.3 & 44.0 & 58.0 \\
\textit{Verb} & 22.5 & 10.6 & \multicolumn{1}{r|}{26.0} & 19.3 & 8.4 & \multicolumn{1}{r|}{25.9} & 20.3 & 8.2 & 34.5 \\
\textit{Adj} & 5.7 & 2.0 & \multicolumn{1}{r|}{19.4} & 10.6 & 3.5 & \multicolumn{1}{r|}{27.5} & 5.7 & 2.0 & 32.1 \\
\textit{Adv} & 8.6 & 7.0 & \multicolumn{1}{r|}{25.6} & 8.6 & 5.1 & \multicolumn{1}{r|}{23.5} & 7.6 & 5.9 & 38.8 \\
\textit{Wh} & 2.1 & 1.4 & \multicolumn{1}{r|}{26.0} & 1.7 & 0.7 & \multicolumn{1}{r|}{21.0} & 2.4 & 1.8 & 44.1 \\
\textit{Func} & 19.2 & 12.1 & \multicolumn{1}{r|}{20.1} & 25.1 & 15.6 & \multicolumn{1}{r|}{24.1} & 17.5 & 10.2 & 30.3 \\
\textit{Stop} & 50.7 & 33.7 & \multicolumn{1}{r|}{25.7} & 50.0 & 30.0 & \multicolumn{1}{r|}{24.8} & 46.8 & 29.5 & 33.7 \\ \hline
\end{tabular}}
\end{table}

Table~\ref{distribution-postag} shows that ER generally increases with WR. This is reasonable, as the more words in one class, the more errors will be generated. However, this does not hold for all PoS classes: the WRs for \textit{Noun}, \textit{Verb} and \textit{Func} are similar, but the ER is much lower for \textit{Verbs}. That is, ASR performs better on verbs than expected based on verb frequency. On IEM, \textit{Verb} has a higher WR than \textit{Noun} and \textit{Func} but much lower ER, and similarly for \textit{Adj} and \textit{Adv} on MOSI. This indicates that in ASR models, some classes of words are more difficult to recognize than others. The \textit{Noun} class is the most difficult to recognize, showing the highest CERs across the corpora. However, other word classes do not have uniform patterns across the corpora. For instance, \textit{Wh} words are the easiest to recognize on MOSI yet the second hardest on MELD. 

The CERs of most word classes increase from IEM to MELD, which is in line with the change of WERs in Table~\ref{wer}. This is plausible as the word error count increases with the WER. Nevertheless, the rate of growth of CER varies from class to class. For example, \textit{Noun} saw a 74\% increase in CER (33.3 to 58.0) compared to 33\% for \textit{Verb} (26.0 to 34.5). Again, this reflects the fact that word classes have different ERs, perhaps because they have different tolerances for speech distortion.

\begin{table}[ht!]
\caption{WR, ER, and CER (all in \%) based on affective score.}
\vspace{-5pt}
\centering
\label{distribution-affect}
\scalebox{0.75}{
\begin{tabular}{lrrrrrrrrr}
\hline
 & \multicolumn{3}{c}{\textbf{IEM}} & \multicolumn{3}{c}{\textbf{MOSI}} & \multicolumn{3}{c}{\textbf{MELD}} \\
 & WR & ER & CER & WR & ER & CER & WR & ER & CER \\ \hline
$V_{low}$ & 2.5 & 3.6 & \multicolumn{1}{r|}{36.1} & 2.3 & 3.6 & \multicolumn{1}{r|}{43.2} & 2.0 & 2.9 & 43.0 \\
$V_{mid}$ & 10.8 & 13.8 & \multicolumn{1}{r|}{29.6} & 9.3 & 11.7 & \multicolumn{1}{r|}{33.9} & 9.8 & 14.2 & 37.7 \\
$V_{high}$ & 13.9 & 16.2 & \multicolumn{1}{r|}{26.3} & 15.6 & 18.9 & \multicolumn{1}{r|}{34.3} & 12.2 & 16.2 & 35.5 \\
$A_{low}$ & 15.0 & 17.4 & \multicolumn{1}{r|}{24.8} & 12.3 & 13.8 & \multicolumn{1}{r|}{30.9} & 13.2 & 17.3 & 35.0 \\
$A_{mid}$ & 11.1 & 14.7 & \multicolumn{1}{r|}{30.7} & 13.6 & 17.5 & \multicolumn{1}{r|}{36.5} & 10.1 & 15.1 & 39.4 \\
$A_{high}$ & 1.0 & 1.4 & \multicolumn{1}{r|}{32.8} & 1.3 & 2.0 & \multicolumn{1}{r|}{43.6} & 0.7 & 1.0 & 40.0 \\
$D_{low}$ & 1.0 & 1.4 & \multicolumn{1}{r|}{28.1} & 1.1 & 1.8 & \multicolumn{1}{r|}{47.2} & 0.8 & 1.2 & 38.9 \\
$D_{mid}$ & 16.7 & 20.5 & \multicolumn{1}{r|}{27.7} & 15.6 & 20.0 & \multicolumn{1}{r|}{36.2} & 15.0 & 20.9 & 36.9 \\
$D_{high}$ & 9.4 & 11.5 & \multicolumn{1}{r|}{27.4} & 10.6 & 11.5 & \multicolumn{1}{r|}{29.2} & 8.1 & 11.2 & 36.7 \\ \hline
\end{tabular}}
\end{table}

Looking at the affective score-based distribution in Table~\ref{distribution-affect} and again see that, in general, the higher the WR, the higher the ER. However, in this case, we don't see classes with similar WRs but very different ERs. Nevertheless, differences still remain: CER decreases as valence and dominance scores increase, i.e., high valence and dominance words are better recognized. In contrast, we see that higher-arousal words are more difficult to recognize. Even so, unlike the classes based on PoS tags, the CERs have a relatively balanced distribution here, and their rates of increase with the WER are at the same level.


We see similar but not identical patterns when we look at the results for individual ASR models. We omit detailed results for brevity but note per-class differences in performance, e.g., \textit{Nouns} are better recognized by KLA (32\% ER) compared to CONF (35\% ER) even though the latter has a lower overall WER. In general, no one model performed the best across all PoS or affective score classes.
 



\subsection{Utterance Length and ASR Performance}
Another factor that is likely to have an effect on ASR performance is utterance length. Generally, longer utterances are more likely to have more word errors compared to shorter ones because there are more opportunities for the ASR model to make errors \cite{sahu19_interspeech}. In fact, \cite{li2022fusing} showed that longer utterances in the transcripts generated by W2V2 have lower WERs compared to shorter ones, hypothesizing that longer utterances contain more contextual information that W2V2 can use to compensate for the speech distortion. However, it is unknown whether other ASR models have the same behavior, so the general impact of utterance length on ASR performance is not yet clear.

To investigate this, we analyze how WER and WR change with utterance length. The four ASR models displayed the same trends, so Table~\ref{wordlength} shows results for W2V2 for brevity. We see that the shorter the utterances, the higher the WER, regardless of corpus. These support the findings of \cite{li2022fusing} and additionally show that if a corpus contains many short utterances, ASR may not work well.

\begin{table}[ht!]
\centering
\caption{WER and word ratio (both in \%) according to different utterance lengths (number of words N).}
\vspace{-5pt}
\label{wordlength}
\scalebox{0.9}{
\begin{tabular}{lrrrrrr}
\hline
 & \multicolumn{2}{c}{\textbf{IEM}} & \multicolumn{2}{c}{\textbf{MOSI}} & \multicolumn{2}{c}{\textbf{MELD}} \\
N & Ratio & WER & Ratio & WER & Ratio & WER \\ \hline
$\le$10 & 59.6 & 46.4 & 53.9 & 40.6 & 71.5 & 73.3 \\
11-20 & 22.9 & 32.2 & 32.2 & 34.5 & 23.5 & 48.6 \\
21-30 & 10.8 & 26.3 & 9.5 & 31.9 & 4.4 & 42.1 \\
$\ge$30 & 6.7 & 25.0 & 4.4 & 32.6 & 0.6 & 38.8 \\ \hline
\end{tabular}}
\vspace{-10pt}
\end{table}


\subsection{ASR Performance on Different Emotions}


It has long been argued that the acoustic characteristics of emotional speech, e.g., prosody variation, can deteriorate ASR performance \cite{li2023exploration}.  Word-level factors, however, are rarely studied. Based on the findings from the above experiments, we conducted a word-level analysis based on different emotions. Specifically, we look at the WER alongside the ratio of \textit{Noun} and short utterance (word count less than 10), which we found to be major factors affecting ASR performance above.

From Table~\ref{emotion-iem}, we can see the WER on Neutral is not the best but instead worse than most of the other emotions in both corpora, which supports the finding in \cite{li2022fusing}. As neutral speech is the least emotionally `distorted', the reason could be that the ratio of short utterances in Neutral is very high, which offsets the benefit of less emotional distortion. We did not see a large discrepancy in the ratio of \textit{Noun} between emotions, although we found that a high ratio harms ASR performance (Sec~\ref{sec:word-class}). 

\begin{table}[ht!]
\centering
\caption{WER, ratio of Noun words and short utterances (all in \%) according to different emotions on IEM and MELD.}
\vspace{-5pt}
\label{emotion-iem}
\scalebox{0.92}{
\begin{tabular}{lcccc}
\hline
\textit{IEM} & \textbf{Ang} & \textbf{Hap} & \textbf{Neu} & \textbf{Sad} \\ \hline
WER & 22.8 & 38.9 & 36.3 & 29.5 \\ 
Noun & 20.0 & 21.3 & 20.4 & 19.5 \\ 
$\le$10 & 44.8 & 55.7 & 60.5 & 52.7 \\ \hline
\end{tabular}}
\end{table}
\vspace{-10pt}
\begin{table}[ht!]
\centering
\scalebox{0.92}{
\begin{tabular}{lccccccc}
\hline
\textit{MELD} & \textbf{Ang} & \textbf{Dis} & \textbf{Fea} & \textbf{Joy} & \textbf{Neu} & \textbf{Sad} & \textbf{Sur} \\ \hline
WER & 52.7 & 53.4 & 58.6 & 59.9 & 58.3 & 52.9 & 65.3  \\ 
Noun & 26.2 & 27.6 & 26.4 & 29.7 & 26.9 & 25.2 & 24.5 \\ 
$\le$10 & 64.1 & 60.1 & 62.1 & 72.4 & 72.8 & 62.0 & 82.8 \\ \hline
\end{tabular}}
\vspace{-10pt}
\end{table}

\subsection{Emotion and ASR Confidence}
Confidence scores can help identify words of high recognition quality for better use of ASR transcripts. \cite{santoso2021speech} and \cite{santoso2022speech} adjusted the attention weights for words using their confidence scores. \cite{pan2020improving} proposed removing words with low confidence and selecting words with the highest confidence from multiple ASR hypotheses \cite{pan2021using}. However, the measurement and use of confidence scores is a complicated task due to out-of-domain words, overconfidence, and so on \cite{qiu2021learning,li2022improving}. Hence, we analyze how confidence varies with emotion by exploring the word confidence scores. We compared two ASR models--CONF and KLA, and also compared the confidence scores generated by KLA with and without calibration. The calibration was applied by using the official procedure in Kaldi\footnote{\url{https://github.com/kaldi-asr/kaldi/blob/master/egs/tedlium/s5/local/confidence_calibration.sh}}. Since we noticed similar patterns of the confidence in all three corpora, we only show the results of IEM for brevity.

\begin{table}[ht!]
\centering
\caption{Distribution of confidence scores of the ASR models.}
\vspace{-5pt}
\label{confidence}
\scalebox{0.85}{
\begin{tabular}{lcccc}
\hline
\multicolumn{1}{l}{} & \textbf{N} & \textbf{CONF} & \textbf{KLA} & \textbf{KLA (calibrated)} \\ \hline
Correct &  & 0.97 & 0.95 & 0.62 \\
Incorrect &  & 0.87 & 0.70 & 0.53 \\ \hline
Ang & \begin{tabular}[c]{@{}c@{}}$\le$10\\ 11-20\\ 21-30\\ $\ge$30\end{tabular} & \begin{tabular}[c]{@{}c@{}}0.96\\ 0.96\\ 0.96\\ 0.96\end{tabular} & \begin{tabular}[c]{@{}c@{}}0.91\\ 0.91\\ 0.90\\ 0.90\end{tabular} & \begin{tabular}[c]{@{}c@{}}0.62\\ 0.61\\ 0.61\\ 0.61\end{tabular} \\ \hline
Hap & \begin{tabular}[c]{@{}c@{}}$\le$10\\ 11-20\\ 21-30\\ $\ge$30\end{tabular} & \begin{tabular}[c]{@{}c@{}}0.92\\ 0.94\\ 0.95\\ 0.94\end{tabular} & \begin{tabular}[c]{@{}c@{}}0.80\\ 0.83\\ 0.86\\ 0.85\end{tabular} & \begin{tabular}[c]{@{}c@{}}0.57\\ 0.58\\ 0.59\\ 0.58\end{tabular} \\ \hline
Neu & \begin{tabular}[c]{@{}c@{}}$\le$10\\ 11-20\\ 21-30\\ $\ge$30\end{tabular} & \begin{tabular}[c]{@{}c@{}}0.91\\ 0.94\\ 0.95\\ 0.96\end{tabular} & \begin{tabular}[c]{@{}c@{}}0.81\\ 0.86\\ 0.88\\ 0.89\end{tabular} & \begin{tabular}[c]{@{}c@{}}0.58\\ 0.59\\ 0.59\\ 0.60\end{tabular} \\ \hline
Sad & \begin{tabular}[c]{@{}c@{}}$\le$10\\ 11-20\\ 21-30\\ $\ge$30\end{tabular} & \begin{tabular}[c]{@{}c@{}}0.92\\ 0.95\\ 0.96\\ 0.97\end{tabular} & \begin{tabular}[c]{@{}c@{}}0.81\\ 0.87\\ 0.89\\ 0.92\end{tabular} & \begin{tabular}[c]{@{}c@{}}0.58\\ 0.59\\ 0.60\\ 0.61\end{tabular} \\ \hline
\end{tabular}}
\vspace{-10pt}
\end{table}

In Table~\ref{confidence}, we noticed that there is a big difference in confidence scores between ASR models. The scores of correctly recognized words in both CONF and KLA are close, but the score for incorrectly recognized words is 0.70 in KLA and 0.87 in CONF. The scores of KLA are smaller than those of CONF in general. Besides, the score increases with utterance length in Ang, Neu, and Sad. The longer the utterance, the higher the confidence, which is in line with the change in WER in Table~\ref{wordlength}. But this trend does not occur in Ang. Note that the score of every emotion increases with utterance length in MELD but barely changes in MOSI, which means using raw confidence scores may not be appropriate. By comparing the confidence with and without the calibration for KLA, we can see that the difference in confidence between correctly and incorrectly recognized words becomes smaller. Similarly, the difference in utterance length almost disappears, which means emotions are more equally treated after calibration. Calibration also adjusts the overconfident behavior of end-to-end ASR models \cite{li2021confidence}. The overall confidence of each emotion is in line with their respective WER in Table~\ref{emotion-iem} that ASR performance is the best on Ang but worst on Hap. These findings demonstrate that confidence scores vary according to different emotions, utterance lengths, and ASR models. This needs to be taken into account if scores are used to adjust word contributions in SER. 

\subsection{The Impact of ASR on SER Performance}
So far, we have investigated emotion corpora to understand the impact of emotion and related word distributions on ASR. Here, we explore the impact of ASR on emotion recognition by computing the SER performance of using ASR transcripts with progressively increased WER (focusing on IEM for brevity).

First, we used all four models to generate four ASR transcripts with their respective WER shown in Table~\ref{wer} and divided every transcript into five parts according to the five sessions required by the 5-fold cross-validation training standard of SER on IEM \cite{li2019improved}. Thus, we obtained four sub-transcripts of each session. Next, for each session, we produced ten sub-transcripts with WER from 5\% to 50\%, with an increase step of 5\%, by randomly replacing a portion of sentences in a sub-transcript (considering different ASR model performances and the speech distortion in real use, we set the highest WER to 50\%). The replacement was done as follows: sentences in transcripts with higher WER replaced those of the same IDs in transcripts with lower WER. If the required WER is outside the range of the WERs of the sub-transcripts, we used ground truths (for low WERs such as 5\%) or worse ASR hypotheses from the CONF model (for high WERs such as 50\%) as replacements to obtain the desired overall WER (we generated 30-best ASR hypotheses using CONF). The proportion of replaced sentences was kept as low as possible in order to preserve the transcripts in their original form as real ASR output.

We trained the SER model on ground truth transcripts from four sessions and tested it on the ASR transcripts from the remaining session (5-fold cross-validation). We used a BERT model \cite{devlin2018bert} to extract textual features. The emotion model comprises two bi-directional LSTM layers (hidden state of 32), a self-attention layer (hidden state of 64 and head number of 16), a dense layer (hidden state of 64) with \textit{ReLU} activation and one output layer with \textit{Softmax} activation. The learning rate is set as 5e-4 with \textit{AdamW} optimizer, the weight decay is set as 1e-5, and the batch size is 64. We trained the models using 150 epochs and reported the best unweighted accuracies.

\vspace{-10pt}
\begin{figure}[ht!]
  \centering
  \includegraphics[width=0.4\textwidth]{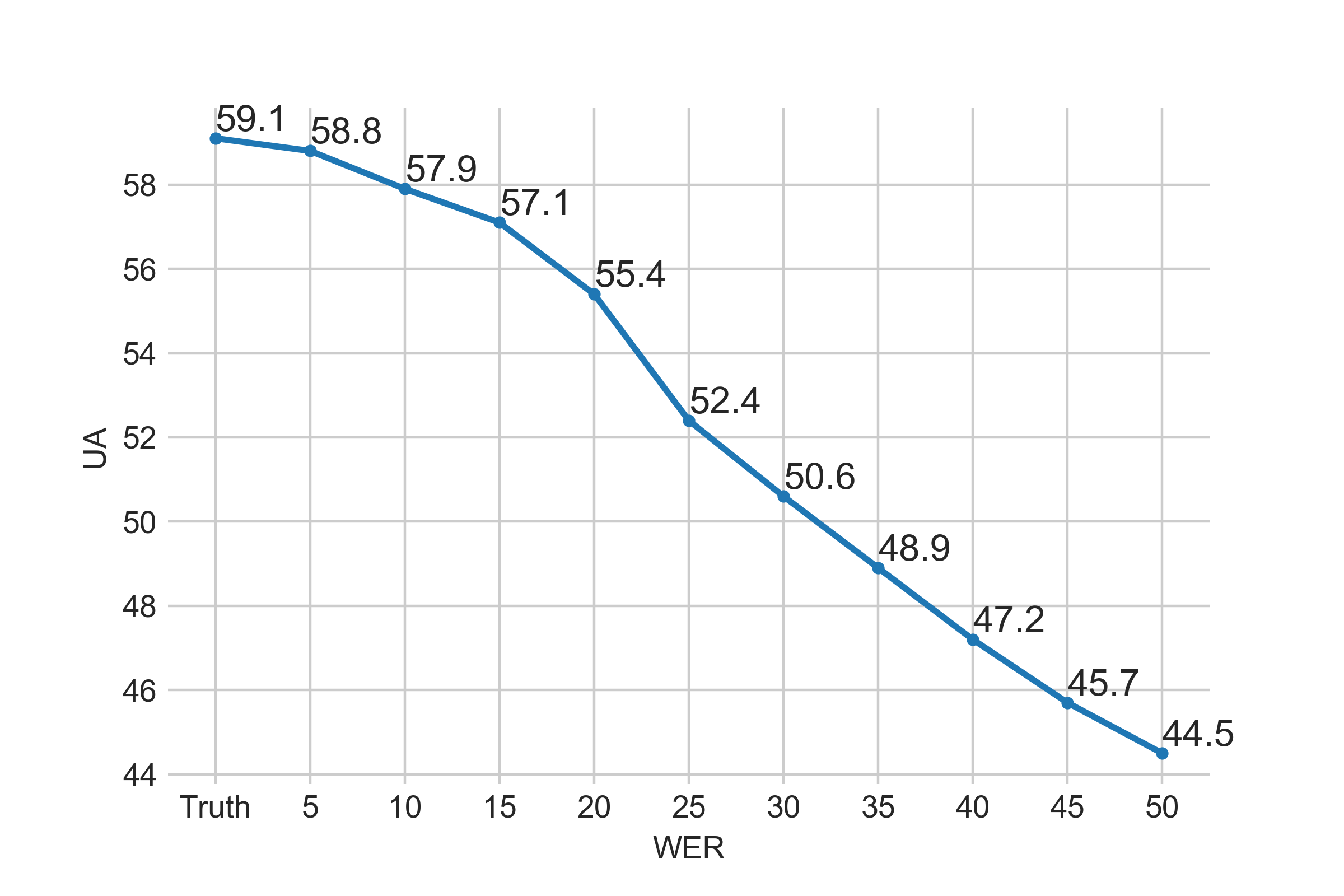}
  \vspace{-5pt}
  \caption{SER accuracy with increasing WER (both in \%).}
  \label{acc}
  \vspace{-5pt}
\end{figure}

Figure~\ref{acc} shows how SER performance varies with WER. We see that WER does affect the SER accuracy. The higher the WER, the lower the accuracy. The accuracy drop is small at first and gradually becomes larger, with a steep drop between 15\% and 25\%. Unlike the finding in \cite{schuller2009emotion} that WER over 30\% only results in an accuracy drop smaller than 3\% on the FAU Aibo Emotion Corpus \cite{batliner2008releasing}, we found that WER can greatly affect SER performance. This may be because we used a different corpus or a different language model (BERT vs. Bag-of-Words). In the future, we will explore how different language models function on imperfect ASR transcripts.

\section{Conclusions}
In this paper, we address a long-existing yet understudied issue: the relationship between ASR and emotion. We demonstrate that word distribution in a corpus is a key factor that affects ASR performance and that differences exist between emotions. The frequencies of different PoS tags, affective scores, and utterance lengths may have a large impact on ASR performance. For example, the more the \textit{Noun} words and short utterances, the higher the WER, regardless of the speaking style (speech collected in the lab, from monologues or TV shows). Moreover, different ASR models behave similarly in these trends, but with minor differences. Some classes of words are easier to recognize for one model but maybe harder for another. Moreover, the confidence scores generated by different models may show large differences, so the use of confidence scores should be carefully examined. Finally, we show that SER accuracy can greatly degrade with high WER. We expect our findings to pave the way for future SER research using imperfect ASR transcripts.

\bibliographystyle{IEEEtran}
\bibliography{mybib}

\begin{thebibliography}{10}
\providecommand{\url}[1]{#1}
\csname url@samestyle\endcsname
\providecommand{\newblock}{\relax}
\providecommand{\bibinfo}[2]{#2}
\providecommand{\BIBentrySTDinterwordspacing}{\spaceskip=0pt\relax}
\providecommand{\BIBentryALTinterwordstretchfactor}{4}
\providecommand{\BIBentryALTinterwordspacing}{\spaceskip=\fontdimen2\font plus
\BIBentryALTinterwordstretchfactor\fontdimen3\font minus
  \fontdimen4\font\relax}
\providecommand{\BIBforeignlanguage}[2]{{%
\expandafter\ifx\csname l@#1\endcsname\relax
\typeout{** WARNING: IEEEtran.bst: No hyphenation pattern has been}%
\typeout{** loaded for the language `#1'. Using the pattern for}%
\typeout{** the default language instead.}%
\else
\language=\csname l@#1\endcsname
\fi
#2}}
\providecommand{\BIBdecl}{\relax}
\BIBdecl

\bibitem{schuller2018speech}
B.~W. Schuller, ``Speech emotion recognition: Two decades in a nutshell,
  benchmarks, and ongoing trends,'' \emph{Communications of the ACM}, vol.~61,
  no.~5, pp. 90--99, 2018.

\bibitem{sebastian2019fusion}
J.~Sebastian, P.~Pierucci \emph{et~al.}, ``{Fusion Techniques for
  Utterance-Level Emotion Recognition Combining Speech and Transcripts},'' in
  \emph{Proceedings of Interspeech 2019}, 2019, pp. 51--55.

\bibitem{yoon2018multimodal}
S.~Yoon, S.~Byun, and K.~Jung, ``Multimodal speech emotion recognition using
  audio and text,'' in \emph{2018 IEEE Spoken Language Technology Workshop
  (SLT)}.\hskip 1em plus 0.5em minus 0.4em\relax IEEE, 2018, pp. 112--118.

\bibitem{li2021feeling}
Y.~Li, ``Feeling estimation device, feeling estimation method, and storage
  medium,'' Aug.~31 2021, {US} Patent 11,107,464.

\bibitem{li2022fusing}
Y.~Li, P.~Bell, and C.~Lai, ``Fusing {ASR} outputs in joint training for speech
  emotion recognition,'' in \emph{ICASSP 2022-2022 IEEE International
  Conference on Acoustics, Speech and Signal Processing (ICASSP)}.\hskip 1em
  plus 0.5em minus 0.4em\relax IEEE, 2022, pp. 7362--7366.

\bibitem{sahu19_interspeech}
S.~Sahu, V.~Mitra, N.~Seneviratne, and C.~Espy-Wilson, ``{Multi-Modal Learning
  for Speech Emotion Recognition: An Analysis and Comparison of ASR Outputs
  with Ground Truth Transcription},'' in \emph{{Proceedings of Interspeech
  2019}}, 2019, pp. 3302--3306.

\bibitem{fernandez2004computational}
R.~Fernandez, ``A computational model for the automatic recognition of affect
  in speech,'' Ph.D. dissertation, Massachusetts Institute of Technology, 2004.

\bibitem{schuller2009emotion}
B.~Schuller, A.~Batliner, S.~Steidl, and D.~Seppi, ``Emotion recognition from
  speech: putting asr in the loop,'' in \emph{2009 IEEE International
  Conference on Acoustics, Speech and Signal Processing}.\hskip 1em plus 0.5em
  minus 0.4em\relax IEEE, 2009, pp. 4585--4588.

\bibitem{santoso2021speech}
J.~Santoso, T.~Yamada, S.~Makino, K.~Ishizuka, and T.~Hiramura, ``Speech
  emotion recognition based on attention weight correction using word-level
  confidence measure.'' in \emph{Interspeech}, 2021, pp. 1947--1951.

\bibitem{busso2008iemocap}
C.~Busso, M.~Bulut, C.-C. Lee, A.~Kazemzadeh, E.~Mower, S.~Kim, J.~N. Chang,
  S.~Lee, and S.~S. Narayanan, ``{IEMOCAP: Interactive emotional dyadic motion
  capture database},'' \emph{{Language Resources and Evaluation}}, vol.~42,
  no.~4, pp. 335--359, 2008.

\bibitem{Zadeh2016}
A.~Zadeh, R.~Zellers, E.~Pincus, and L.-P. Morency, ``Multimodal sentiment
  intensity analysis in videos: Facial gestures and verbal messages,''
  \emph{IEEE Intelligent Systems}, vol.~31, no.~6, pp. 82--88, 2016.

\bibitem{poria2019meld}
S.~Poria, D.~Hazarika, N.~Majumder, G.~Naik, E.~Cambria, and R.~Mihalcea,
  ``Meld: A multimodal multi-party dataset for emotion recognition in
  conversations,'' in \emph{Proceedings of the 57th Annual Meeting of the
  Association for Computational Linguistics}, 2019, pp. 527--536.

\bibitem{li2019improved}
Y.~Li, T.~Zhao, and T.~Kawahara, ``{Improved End-to-End Speech Emotion
  Recognition Using Self Attention Mechanism and Multitask Learning},'' in
  \emph{Proceedings of Interspeech 2019}, 2019, pp. 2803--2807.

\bibitem{watanabe2018espnet}
S.~Watanabe, T.~Hori, S.~Karita, T.~Hayashi, J.~Nishitoba, Y.~Unno, N.-E.~Y.
  Soplin, J.~Heymann, M.~Wiesner, N.~Chen \emph{et~al.}, ``Espnet: End-to-end
  speech processing toolkit,'' \emph{Proc. Interspeech 2018}, pp. 2207--2211,
  2018.

\bibitem{marcus1993building}
M.~Marcus, B.~Santorini, and M.~A. Marcinkiewicz, ``Building a large annotated
  corpus of english: The penn treebank,'' \emph{Computational Linguistics},
  vol.~19, no.~2, pp. 313--330, 1993.

\bibitem{bird2006nltk}
S.~Bird, ``{NLTK}: the natural language toolkit,'' in \emph{Proceedings of the
  COLING/ACL 2006 Interactive Presentation Sessions}, 2006, pp. 69--72.

\bibitem{warriner2013norms}
A.~B. Warriner, V.~Kuperman, and M.~Brysbaert, ``Norms of valence, arousal, and
  dominance for 13,915 english lemmas,'' \emph{Behavior research methods},
  vol.~45, no.~4, pp. 1191--1207, 2013.

\bibitem{li2023exploration}
Y.~Li, Y.~Mohamied, P.~Bell, and C.~Lai, ``Exploration of a self-supervised
  speech model: A study on emotional corpora,'' in \emph{2022 IEEE Spoken
  Language Technology Workshop (SLT)}.\hskip 1em plus 0.5em minus 0.4em\relax
  IEEE, 2023, pp. 868--875.

\bibitem{santoso2022speech}
J.~Santoso, T.~Yamada, K.~Ishizuka, T.~Hashimoto, and S.~Makino, ``Speech
  emotion recognition based on self-attention weight correction for acoustic
  and text features,'' \emph{IEEE Access}, vol.~10, pp. 115\,732--115\,743,
  2022.

\bibitem{pan2020improving}
Y.~Pan, B.~Mirheidari, M.~Reuber, A.~Venneri, D.~Blackburn, and H.~Christensen,
  ``Improving detection of alzheimer’s disease using automatic speech
  recognition to identify high-quality segments for more robust feature
  extraction,'' in \emph{Proceedings of Interspeech 2020}.\hskip 1em plus 0.5em
  minus 0.4em\relax International Speech Communication Association (ISCA),
  2020, pp. 4961--4965.

\bibitem{pan2021using}
Y.~Pan, B.~Mirheidari, J.~M. Harris, J.~C. Thompson, M.~Jones, J.~S. Snowden,
  D.~Blackburn, and H.~Christensen, ``Using the outputs of different automatic
  speech recognition paradigms for acoustic-and bert-based alzheimer's dementia
  detection through spontaneous speech.'' in \emph{Interspeech}, 2021, pp.
  3810--3814.

\bibitem{qiu2021learning}
D.~Qiu, Q.~Li, Y.~He, Y.~Zhang, B.~Li, L.~Cao, R.~Prabhavalkar, D.~Bhatia,
  W.~Li, K.~Hu \emph{et~al.}, ``Learning word-level confidence for subword
  end-to-end asr,'' in \emph{ICASSP 2021-2021 IEEE International Conference on
  Acoustics, Speech and Signal Processing (ICASSP)}.\hskip 1em plus 0.5em minus
  0.4em\relax IEEE, 2021, pp. 6393--6397.

\bibitem{li2022improving}
Q.~Li, Y.~Zhang, D.~Qiu, Y.~He, L.~Cao, and P.~C. Woodland, ``Improving
  confidence estimation on out-of-domain data for end-to-end speech
  recognition,'' in \emph{ICASSP 2022-2022 IEEE International Conference on
  Acoustics, Speech and Signal Processing (ICASSP)}.\hskip 1em plus 0.5em minus
  0.4em\relax IEEE, 2022, pp. 6537--6541.

\bibitem{li2021confidence}
Q.~Li, D.~Qiu, Y.~Zhang, B.~Li, Y.~He, P.~C. Woodland, L.~Cao, and T.~Strohman,
  ``Confidence estimation for attention-based sequence-to-sequence models for
  speech recognition,'' in \emph{ICASSP 2021-2021 IEEE International Conference
  on Acoustics, Speech and Signal Processing (ICASSP)}.\hskip 1em plus 0.5em
  minus 0.4em\relax IEEE, 2021, pp. 6388--6392.

\bibitem{devlin2018bert}
J.~Devlin, M.-W. Chang, K.~Lee, and K.~Toutanova, ``{BERT: Pre-training of deep
  bidirectional transformers for language understanding},'' \emph{Proceedings
  of NAACL-HLT 2019}, pp. 4171--4186, 2019.

\bibitem{batliner2008releasing}
A.~Batliner, S.~Steidl, and E.~N{\"o}th, ``Releasing a thoroughly annotated and
  processed spontaneous emotional database: the fau aibo emotion corpus,'' in
  \emph{Programme of the Workshop on Corpora for Research on Emotion and
  Affect}, 2008, p.~28.

\end{thebibliography}

\end{document}